\documentclass[aps,prb,superscriptaddress,draft,showpacs,intlimits,amsmath,amssymb,floats,floatfix,twocolumn]{revtex4}
\usepackage[final]{graphicx}
\usepackage{amsmath}
\begin{document}
\title{Quantitative comparison of single- and two-particle properties in the cuprates}
\author{W. Prestel}
\affiliation{Walther Meissner Institut, Bayerische Akademie der
Wissenschaften, 85748 Garching, Germany}
\author{F. Venturini}
\altaffiliation[Permanent address: ]{Mettler-Toledo (Schweiz)
GmbH, 8606 Greifensee, Switzerland}
\affiliation{Walther Meissner Institut, Bayerische Akademie der
Wissenschaften, 85748 Garching, Germany}
\author{B. Muschler}
\affiliation{Walther Meissner Institut, Bayerische Akademie der
Wissenschaften, 85748 Garching, Germany}
\author{I. T\"utt\H{o}}
\affiliation{Research Institute for Solid State Physics and Optics,
Hungarian Academy of Sciences, 1525 Budapest, Hungary}
\author{R. Hackl}
\email{hackl@wmi.badw.de}
\affiliation{Walther Meissner Institut, Bayerische Akademie der
Wissenschaften, 85748 Garching, Germany}
\author{M. Lambacher}
\affiliation{Walther Meissner Institut, Bayerische Akademie der
Wissenschaften, 85748 Garching, Germany}
\author{A. Erb}
\affiliation{Walther Meissner Institut, Bayerische Akademie der
Wissenschaften, 85748 Garching, Germany}
\author{Seiki Komiya}
\affiliation{Central Research Institute of the Electric Power 
Industry, Komae, Tokyo 201-8511, Japan}
\author{Shimpei Ono}
\affiliation{Central Research Institute of the Electric Power 
Industry, Komae, Tokyo 201-8511, Japan}
\author{Yoichi Ando}
\affiliation{Institute of Scientific and Industrial Research,
Osaka University, Ibaraki, Osaka 567-0047, Japan}
\author{D. Inosov}
\altaffiliation[Present address: ]{MPI f\"ur 
Festk\"orperforschung, Heisenbergstr. 1, 70569 Stuttgart, Germany}
\affiliation{IFW Dresden, P.O. Box 270116, 01171 Dresden, Germany}
\author{V.\,B. Zabolotnyy}
\affiliation{IFW Dresden, P.O. Box 270116, 01171 Dresden, Germany}
\author{S.\,V. Borisenko}
\affiliation{IFW Dresden, P.O. Box 270116, 01171 Dresden, Germany}
\begin{abstract}
We explore the strong variations of the electronic properties of copper-oxygen compounds across the doping phase diagram in a quantitative way. To this end we calculate the electronic Raman response on the basis of results from angle-resolved photoemission spectroscopy (ARPES). In the limits of our approximations we find agreement on the overdoped side and pronounced discrepancies at lower doping. In contrast to the successful approach for the transport properties at low energies, the Raman and the ARPES data cannot be reconciled by adding angle-dependent momentum scattering. We discuss possible routes towards an explanation of the suppression of spectral weight close to the $(\pi,0)$ points which sets in abruptly close to 21\% doping.
\end{abstract}
\maketitle

\section{Introduction}
\label{sec:1}
Superconductivity in the copper-oxygen compounds (cuprates) \cite{Bednorz:1986} occurs at transition temperatures $T_c$ above 100~K with the origin still controversial. On a basic level the problem can be cast into the question as to whether the condensed state at low-temperature is an instability of the normal metal as in conventional systems or a new ground state. If superconductivity results from interactions between the electrons via an intermediary boson the theoretical approach proposed by
Eliashberg \cite{Eliashberg:1960,Carbotte:1990,Carbotte:2003} should provide an
at least qualitative description of the condensation of electrons into Cooper pairs. Signatures of
the coupling boson should also determine the
properties above $T_c$. Traditionally, one looks at
renormalization effects in the single-particle or transport
responses as a function of temperature, specifically around $T_c$
where abrupt changes are to be expected
\cite{Engelsberg:1963,Allen:1972,Cuk:2005}. If the spectral shape of the coupling function can be retrieved \cite{McMillan:1965,Carbotte:1999,Kordyuk:2010} one may be in a position to derive the relevant interaction(s). However, the results in the cuprates are
far from converging into a unified picture. The unsuccessful
search for clear signatures of a dominating retarded interaction
was one reason why new ground states were and are being studied
intensively \cite{Anderson:1987,Lee:2006,Anderson:2007}.

There were early proposals that the superconductivity in the
cuprates could be a property of the electrons alone. A prominent
example is the resonating valence bond (RVB) state with
antiferromagnetic coupling of nearest neighbor spins
\cite{Anderson:1987} which emerges from the N\'eel state upon
doping \cite{Lee:2006}. From an experimental point of view this
proposal is much harder to pin down than superconductivity as an
instability or a small perturbation of the normal state. There is
indeed an arsenal of methods to analyze data in terms of an
Eliashberg-type strong coupling approach
\cite{Carbotte:1990,Cuk:2005,McMillan:1965,Carbotte:1999,Kordyuk:2010,Dahm:2009} while ideas for
experiments unveiling new ground states are
scarce.

Our Raman experiments aim at disentangling strong coupling effects
evolving mainly as a function of temperature \cite{Carbotte:1999}
from phase transitions and cross-over phenomena occurring as a
function of doping (see also Ref.~\cite{Muschler:2010}). We develop an analysis using a phenomenological
description of the response \cite{Venturini:2003}. Starting from single-particle
properties observed by angle-resolved photoemission spectroscopy
(ARPES) we calculate the electronic Raman response (ERS) in both
the normal and the superconducting states. While ARPES measures
occupied single-electron states ERS yields - similarly as, e.g.,
optical transport (IR) - a weighted convolution of occupied and
empty states. Therefore the two-particle spectra contain
additional information originating from the interaction between
the hole and the electron created in the scattering process.

\section{Experiment and theoretical background}
\label{sec:2}
In the Raman process an electron is scattered through an
intermediate high-energy state into an unoccupied level (for a more detailed discussion see Refs.~\cite{Muschler:2010,Devereaux:2007}). The
energy transferred to the electron corresponds to the difference
between the incoming and the outgoing photon,
$\Omega=\omega_i-\omega_s$. As a result of the Coulomb interaction
only charge fluctuations inside the unit cell can be observed. The
resulting selection rules lead to form factors $\gamma^{\mu}_{\bf k}$
weighing out symmetry-specific regions of the Brillouin zone. For
crossed photon polarizations aligned along the principle axes (of the CuO$_2$ plane) the
$B_{2g}$ symmetry is projected out and $\gamma_{\bf k}^{B2g}
\propto \sin k_x\sin k_y$ . In $B_{1g}$ symmetry with both
polarizations rotated by $45^{\circ}$ the form factor reads
$\gamma_{\bf k}^{B1g} \propto \cos k_x - \cos k_y$ as shown on the r.h.s. of
Fig.~\ref{fig:lsco-T}. Excitations with $A_{1g}$ symmetry will not be considered here.\\

For the $B$ symmetries the Raman response function $\chi^{\prime\prime}_{\gamma\gamma}(\Omega)$ can be expressed directly in terms of a generalized susceptibility
$\tilde{\chi}_{a,b}(\Omega)$ which, in Nambu representation,
reads \cite{Devereaux:2007}
\begin{equation}
		\begin{split}
  		\tilde{\chi}_{a,b}(i\Omega)=\\
  		\frac{T}{N}\sum_{{\bf k},\sigma}\sum_{m}Tr[\hat{a}({\bf k}) 
  		\hat{G}({\bf k},i\omega_m)\hat{b}({\bf k})\hat{G}({\bf k},i\omega_m + i\Omega)]
  	\end{split}
  \label{eq:chi}
\end{equation}
with $T$ the temperature, $N$ the number of {\bf k}-points $\hat{a},~\hat{b}$ the vertices and $\hat{G}({\bf k},i\omega)$ the matrix Green function. The vertices can be expressed in terms of Pauli matrices $\hat{\tau}_i$. For charge excitations, $\hat{a}({\bf k}) = \hat{\tau}_3 a({\bf k})$. $a({\bf k})$ can be either unity or one of the vertices $\gamma_{\bf k}^{\mu}$ or the electron velocity $v_{\bf k}$ to yield the density or the Raman response or, respectively, the conductivity. In general, the vertex $\hat{b}$ must be renormalized \cite{Devereaux:2007} but in the approximation here we use only bare vertices. $G_{1,1}^{\prime\prime}({\bf k},\omega)=-\pi A({\bf k},\omega)$ is the
imaginary part of the renormalized Green function which is proportional to the spectral function $A({\bf k},\omega)$ which, for the occupied states, can be measured by ARPES.

In strong-coupling theory the components of $\hat{G}({\bf k},\omega)$ read \cite{Cuk:2005,Mahan:2000}
\begin{align}
  G_{1,1}({\bf k},\omega) &=&
  \frac{\omega Z({\bf k},\omega)+\xi_{\bf k}+\chi({\bf k},\omega)}{[\omega Z({\bf k},\omega)]^2-[\xi_{\bf k}+\chi({\bf k},\omega)]^2-[\Phi({\bf k},\omega)]^2}\\
  G_{1,2}({\bf k},\omega) &=&
  \frac{-\Phi({\bf k},\omega)}{[\omega Z({\bf k},\omega)]^2-[\xi_{\bf k}+\chi({\bf k},\omega)]^2-[\Phi({\bf k},\omega)]^2}
  \label{eq:G}
\end{align}
etc. with $\xi_{\bf k} = \epsilon_{\bf k}-\mu$ the bare band structure. The complex functions $Z({\bf k},\omega)$, $\chi({\bf k},\omega)$, and $\Phi({\bf k},\omega)$ must be found self consistently and  describe all interactions of the electrons. In the weak coupling limit $Z=1$, $\chi=0$, and $\Phi=\Delta_{\bf k}$. For $Z \ne 1$ and $\chi$ finite the usual self energy $\Sigma = \Sigma^{\prime} + i\Sigma^{\prime\prime}$ is related to $Z$ and $\chi$ as $\Sigma^{\prime\prime} = -\omega Z^{\prime\prime} +\chi^{\prime\prime}$ \cite{Cuk:2005,Eschrig:2003,Chubukov:2004}, $Z=1-\Sigma/\omega$ and $Z=(1+\Sigma^{\prime}/\omega)^{-1} $below and above $T_c$, respectively, and $\Phi = Z\Delta_{\bf k}$ \cite{Eschrig:2003} in lowest order approximation. Since $\Sigma^{\prime}$ is anti-symmetric and $\Sigma^{\prime\prime}$ vanishes faster than $\omega$ below $T_c$ even in the presence of impurities or nodes, $Z$ is always defined. In the normal state $\Phi$ vanishes and $G^{-1}_{1,1} \equiv G^{-1}_0-\Sigma$.

For the calculation of
$\chi^{\prime\prime}_{\gamma\gamma}(\Omega)$  we need both the occupied
and the empty states and model functions for
$A({\bf k},\omega)$ must be derived from the experimental ARPES
spectra $A_{\exp}({\bf k},\omega)$ being cut off at the chemical
potential $\mu(T)$ by the Fermi function $f(T,\omega)$. Ideally, one has to microscopically derive
expressions for the bare band structure $\xi_{\bf k}=\epsilon_{\bf
k}-\mu$ and for ${\Sigma}$, $Z$, $\chi$, and $\Phi$ in order to obtain a spectral function
$A({\bf k},\omega)$ which, after multiplication with the matrix
element $|M_{f,i}|^2$ and convolution with the experimental resolution function
$R_{\exp}$, reproduces $A_{\exp}=(f\,A|M_{f,i}|^2)\otimes R_{\exp}$.

Since we do not focus on microscopic models here neither for the normal nor for the superconducting state we use phenomenological approximations for both the self energy and the band structure which reproduce the observed ARPES data satisfactorily. A tight binding description with nearest and next-nearest neighbor hopping
\begin{equation}
   \xi_{\bf k} = - 2t(\cos k_x+\cos k_y) + 4t^{\prime}(\cos k_x\cos k_y) - \mu
   \label{eq:xi}
\end{equation}
using $t=0.25$\,eV for the nearest neighbor hopping integral and $t^{\prime}/t=0.35$ for ${\rm La_{2-x}Sr_xCuO_4}$ (LSCO), $t^{\prime}/t=0.40$ for  ${\rm Bi_2Sr_2CaCu_2O_{8+\delta}}$ (Bi-2212), and a self energy of the form
\begin{equation}
   \Sigma^{\prime\prime}=-\left[\sqrt{(\alpha\omega)^2+(\beta T)^2+c_0^2}+c_{\bf k}\right],
   \label{eq:Sigma}
\end{equation}
which is inspired by (but not directly drivable from) the marginal Fermi-liquid (mFL) approach, $\Sigma^{\prime\prime}=-\max(\alpha|\omega|,\beta T)$ \cite{Varma:1989}, yield good agreement with photoemission. The chemical potential $\mu(T)$ is adjusted to give the correct filling and a momentum dependent elastic scattering term $c_{\bf k}$ is additionally included. For $c_0=0$, $\Sigma^{\prime\prime}$ is strictly linear at $T=0$. With $c_0$ finite $\Sigma^{\prime\prime}$ varies always quadratically at low energy. We assume that $\Sigma^{\prime\prime}$ saturates above an energy $\omega_0$ which approximately corresponds to the Joffe-Regel limit equivalent to a mean free path $\ell \approx a$ with $a$ the lattice constant.  The real part of the self energy $\Sigma^{\prime}$ is obtained from Eq.~(\ref{eq:Sigma}) by Kramers-Kr\"onig transformation. In the superconducting state we follow the phenomenology of Inosov \textit{et al.} \cite{Inosov:2007}.

\section{Results and discussion}
\label{sec:3}

We have measured electronic Raman spectra of Bi-2212 and LSCO above and below the ``critical'' doping $p=x \approx 0.21$ at which the $B_{1g}$ response changes qualitatively. In Bi-2212 we have evidence that the transition is quite abrupt \cite{Venturini:2002b,Blanc:2009}. Neither in ${\rm Tl_2Ba_2CaCuO_{6+\delta}}$ (Tl-2201) nor in LSCO  the doping levels are sufficiently dense to allow this conclusion. However, the crossover in the range of 21\% is clearly seen and seems to be one of the generic features of the Raman response in the cuprates \cite{Muschler:2010}. We now try to answer the question in which way this feature can be related to other properties of the cuprates as observed for example in ARPES or transport experiments.
%%%%%%%%%%%%%%%%%%%%%%%% FIGURE 1: LSCO T-dependence %%%%%%%%%%%%%%%%%%%%%%%%%%%%%%%%
\begin{figure*}
  \centering
  \resizebox{1.7\columnwidth}{!}{\includegraphics{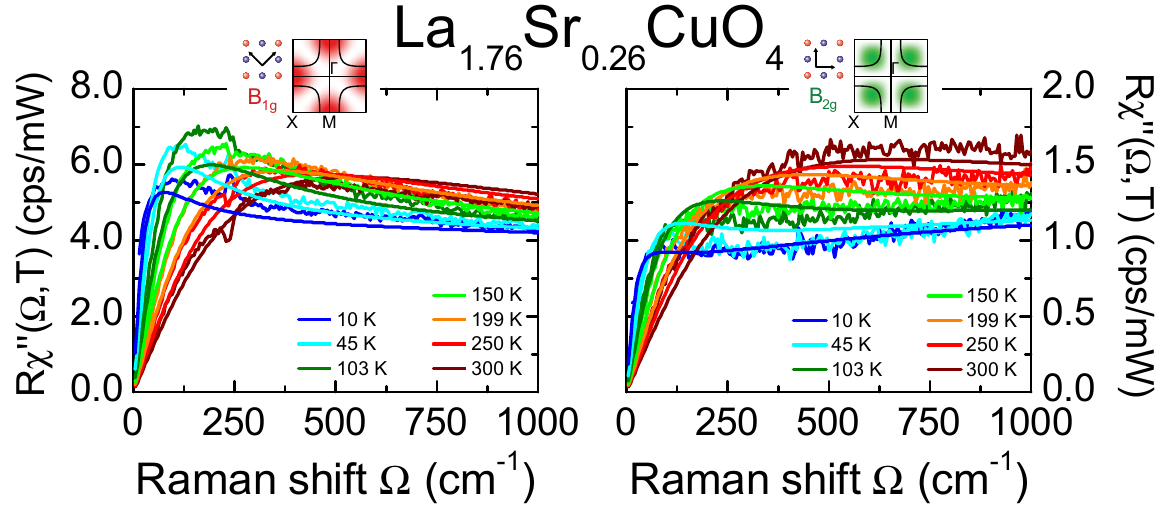}}
  \caption{Temperature dependence of the electronic Raman response of
  overdoped non-superconducting ${\rm La_{1.74}Sr_{0.26}CuO_4}$.
  The smooth lines represent the phenomenology described in the text and
  are based on an analytic approximation to the photoemission results.}
  \label{fig:lsco-T}
\end{figure*}
%%%%%%%%%%%%%%%%%%%%%%%%%%%%%%%%%%%%%%%%%%%%%%%%%%%%%%%%%%%%%%%%%%%%%%%%%%%%%%%%%%%

We start with non-superconducting LSCO ($x=0.26$) and parameterize the self-energy as $\alpha=1.1$, $\beta=2.5$, $c_0/t=0.04$, $c_{\bf k}=0$, and $\mu(100\,K)/t=1.3$ to obtain the proper filling and satisfactory
agreement with the ARPES spectra \cite{Damascelli:2003,Yoshida:2006,JChang:2008}. In addition, the temperature
dependence of the resistivity is well reproduced. With these parameters, optimized for
${\rm La_{1.74}Sr_{0.26}CuO_4}$,  and using Eq.~(\ref{eq:chi}) we
arrive at the Raman spectra shown in Fig.~\ref{fig:lsco-T}. The
agreement is remarkable since only a single intensity is set at
150~K. Symmetry and temperature dependence follow then within the
model.

It has been shown before that the longitudinal and the Hall
resistivity can be reproduced in a Boltzmann approach with similar
assumptions \cite{Nakamae:2003}. Here, we use full {\bf k} sums to
calculate the spectra at finite energies. The limitation to the
Fermi surface would suppress essential structures of the spectra
originating from the anisotropy of $\xi_{\bf k}$ and the proximity of the van
Hove singularity to $\mu$. The agreement reaches beyond the energy
range shown here and leaves only little spectral weight
unexplained up to 1~eV. The agreement between various single- and two-particle probes at
high doping demonstrates that the self energy alone captures the
essential many-body physics and that the lowest order approximation given in
Eq.~(\ref{eq:chi}) is sufficient for the calculation of response
functions.

The only type of microscopic models for $\Sigma$ which qualitatively reproduce the variation with
energy and temperature compatible with Eq.~(\ref{eq:Sigma}) are based on fluctuations. They lead to ``marginal'' behavior with strongly reduced quasiparticle weight, $Z \rightarrow 0$, at the Fermi surface, in certain doping and temperature ranges. The marginal Fermi-liquid model (mFL) \cite{Varma:1989} was the first proposal which predicted
expressions similar to Eq.~(\ref{eq:Sigma}). Later, circulating orbital currents were proposed to be the microscopic origin of this phenomenology \cite{Varma:1997} which may leave an imprint on the  Bragg peaks in neutron scattering experiments \cite{YLi:2008}. Fluctuations of an incipient charge density wave (CDW) \cite{Caprara:1999,Grilli:2009} or a Fermi surface deformation \cite{Metzner:2003} are also candidates. They lead to Fermi
liquid like variations of $\Sigma$ at low temperature, actually well below $T_c$, and predict a strong momentum dependence of the self energy close to the quantum critical point where the transition temperature to the ordered or partially ordered state approaches or extrapolates to zero. In fact, indications of CDW fluctuations, nematic or even long-ranged ordering have been observed in various experiments \cite{Tranquada:1995,Tranquada:2004,Klauss:2000,Fink:2009,Hinkov:2010,Vojta:2010}.

%%%%%%%%%%%%%%%%%%%%%%%% FIGURE 2: Bi-2212&LSCO--doping %%%%%%%%%%%%%%%%%%%%%%%%%%%
\begin{figure*}
  \centering
  \resizebox{1.25\columnwidth}{!}{\includegraphics{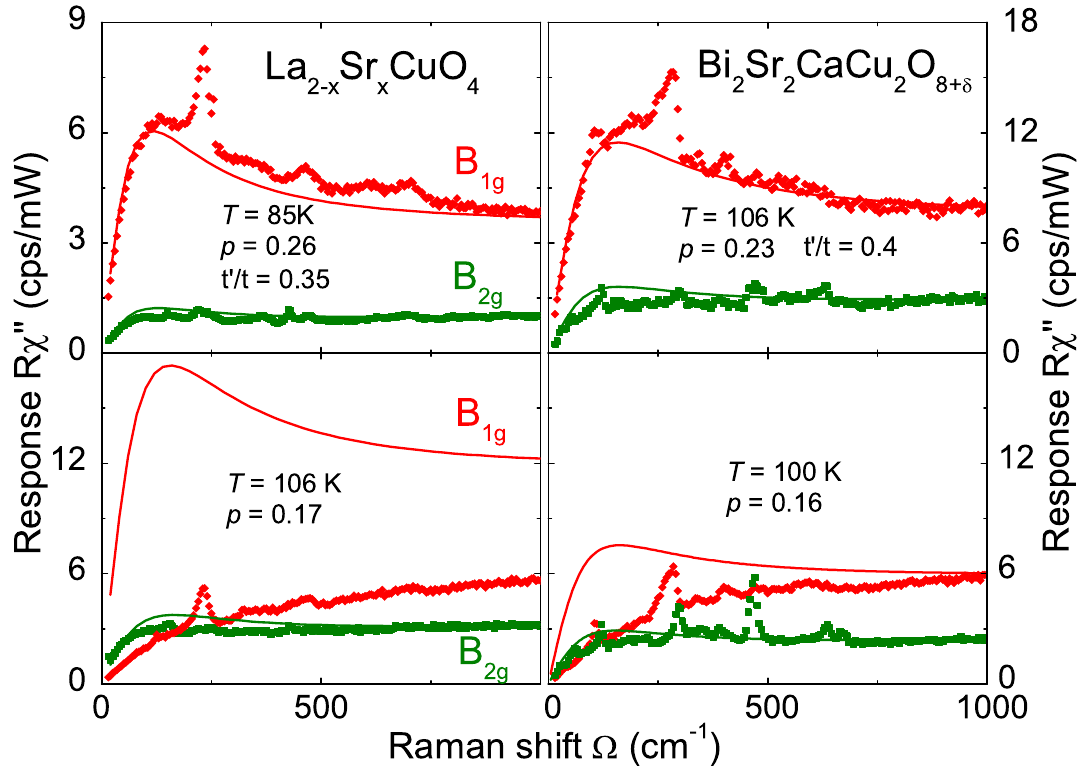}}
  \caption{Doping dependence of the electronic Raman response of
  $\rm Bi_2Sr_2CaCu_2O_{8+\delta}$ and ${\rm La_{2-x}Sr_{x}CuO_4}$.
  There is little variation for the $B_{2g}$ spectra. The $B_{1g}$ responses
  change abruptly at $p \simeq 0.21$ and cannot be described any more by spectral
  functions derived from the ARPES results in lowest order approximation [Eq.~(\ref{eq:chi})].}
  \label{fig:Bi-LSCO}
\end{figure*}
%%%%%%%%%%%%%%%%%%%%%%%%%%%%%%%%%%%%%%%%%%%%%%%%%%%%%%%%%%%%%%%%%%%%%%%%%%%%%%%%%%%

At high doping, $p > 0.21$, an anisotropy of the self energy is not observed. Upon decreasing the doping level the ARPES spectra show only little variation and even around optimal doping there are
well defined quasiparticle peaks on the entire Fermi
surface. All changes appear to be continuous \cite{Damascelli:2003,Yoshida:2006,JChang:2008}. In
contrast, the $B_{1g}$ Raman spectra change abruptly close to $p =
0.21$ \cite{Venturini:2002b,Blanc:2009}. Since there is no discontinuity in the
ARPES spectra around $p = 0.21$ we can use the same model for
$\Sigma$ and $\xi_{\bf k}$ on either side of the crossover. The
comparison between the prediction on the basis of the ARPES
results and the observed $B_{1g}$ Raman spectra for samples above
and below $p = 0.21$ are shown in Fig.~\ref{fig:Bi-LSCO}. In
$B_{2g}$ symmetry, the spectra are well described at both doping
levels and, beyond that, down to the lowest carrier concentration inside the
superconducting dome \cite{Muschler:2010}. The $B_{1g}$ spectra, however, drop considerably
below the simulation in the range below 1000~cm$^{-1}$. While the
spectral shapes are similar in Bi-2212 and LSCO the overall
intensity in LSCO is subject to variations of the cross section
due to resonance effects.

The main result here is twofold. (i) A strong interaction sets in below approximately 21\% doping which (ii) manifests itself predominantly in the two particle properties. Indications of this interaction have been seen a long time ago by nuclear magnetic resonance (NMR) \cite{Alloul:1989} which measures a similar response function [see Eq.~(\ref{eq:chi})]. With Raman scattering we can additionally pin down the momentum dependence \cite{Venturini:2002b}. In IR experiments anomalies are mainly observed in the $c$-axis conductivity of $R\rm Ba_2Cu_3O_{6+x}$ ($R$-123 with $R=$\,Y,\,Nd,\,La) compounds \cite{Holmes:1993,Bernhard:1998,Basov:2005} while the transport in the $a-b$ plane is essentially captured by a Fermi liquid or mFL phenomenology \cite{vdMarel:2003,Hussey:2003,Cooper:2009}. This is not unexpected and compatible with the Raman results when the BZ projections of the respective methods are taken into account. Then, due to the band structure of the cuprates \cite{Andersen:1995} in-plane IR and $B_{2g}$ Raman project predominantly the nodes while out-of-plane transport and $B_{1g}$ Raman are more sensitive at the anti-node \cite{Devereaux:2003}.
%%%%%%%%%%%%%%%%%%%%%%%%%%%%%%%%%%%%%%%%%%%%%%%%%%%
\begin{figure*}
  \centering
  \resizebox{1.35\columnwidth}{!}{\includegraphics{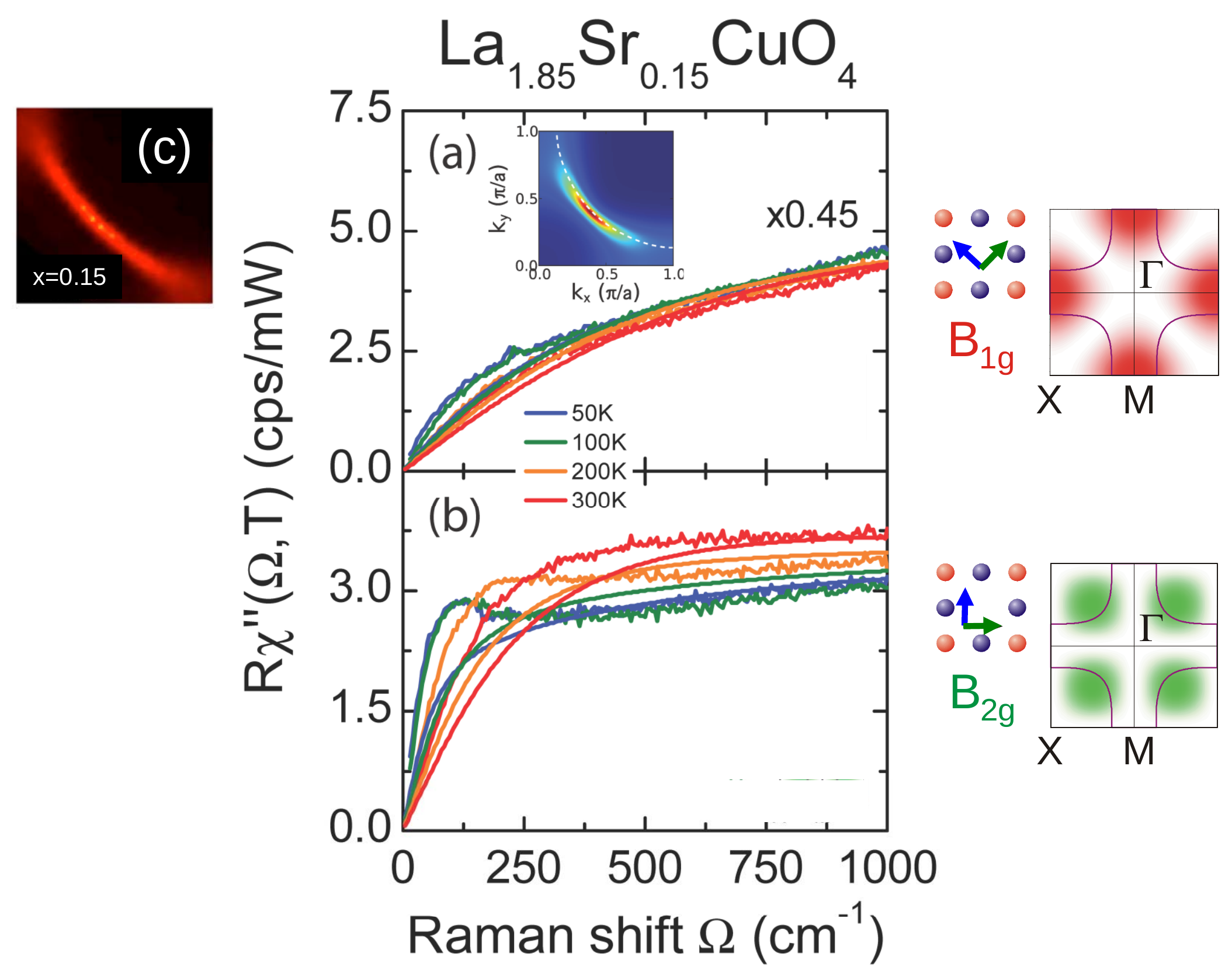}}
  \caption{Temperature dependence of the electronic Raman response of
  optimally doped ${\rm La_{1.85}Sr_{0.15}CuO_4}$.
  The smooth lines represent the calculations. Here, an additional anisotropic contribution from elastic collisions according to Eq.~(\ref{eq:ck}) with $c_0=c_1=0$ and $c_2=0.1t$. The inset in (a) shows the spectral function at the Fermi energy [$A({\bf k},\omega=0)$] in false-color representation (red high, blue low intensity) for these parameters. The suppression of intensity around ($\pi,0$) is much stronger than experimentally observed at this doping level as shown in panel (c) (From Ref.~\cite{Yoshida:2006} with permission.)
  }
  \label{fig:imp}
\end{figure*}
%%%%%%%%%%%%%%%%%%%%%%%%%%%%%%%%%%%%%%%%%%%%%%%%%%%%%%%%%%%%%%%%%%%%%%%%%%%%%%%%%%

It has been suggested that an increasing contribution from anisotropic elastic scattering of electrons leads to the dichotomy between nodal ($\pi/2,\pi/2$) and anti-nodal ($\pi,0$) particles \cite{Cooper:2009,Kampf:1997,Abrahams:2000}. We check this proposal here by adding an energy and temperature independent anisotropic scattering rate varying as
\begin{equation}
   c_{\bf k} = c_1 +\frac{c_2}{2}(\cos k_x-\cos k_y)^2.
   \label{eq:ck}
\end{equation}
With $c_1$ and $c_2$ properly selected the variation of the scattering rate on the Fermi surface as used for Tl-2201 \cite{Hussey:2003,AbdelJawad:2007} (but not any more for LSCO \cite{Cooper:2009}) is reproduced to within a few percent. In Fig.~\ref{fig:imp} we compare the resulting Raman spectra with the experimental data of LSCO at optimal doping. For our choice of parameters the phenomenological curves fit the shape and the temperature dependence of the $B_{1g}$ data reasonably well but deviate now significantly from those in $B_{2g}$ symmetry. In LSCO a fit to the $B_{1g}$ results is still possible since the temperature dependence of the spectra remains metallic. In contrast, the spectra in Bi-2212 exhibit even a slightly insulating variation with temperature around optimal doping \cite{Venturini:2002b} and agreement between the calculated and the experimental $B_{1g}$ spectra cannot be obtained any further. We conclude that an elastic term is insufficient to reconcile single- and two-particle properties at optimal doping in the same approximation as at high doping and that dynamic interactions are at the origin of the observed effects. In addition, it is not completely surprising that higher order corrections may become necessary when entering the low doping range.

The reason why the transport data can be reproduced reasonably well in terms of an isotropic  Boltzmann approximation \cite{Cooper:2009} is not immediately obvious. The linear dispersion, $\epsilon_{\bf k} \approx v_{\bf k}({\bf p}-{\bf p}_F)$, around the Fermi momentum ${\bf p}_F$ and the specific variation of the current vertices used in that approach may be part of the explanation. In earlier studies of the superconducting  response we actually found substantial differences  between Fermi surface and {\bf k} integration. After all, the transport is mainly due to nodal quasiparticles which survive down to at least $p=0.05$ and may be mostly blind for the effects we find around $(\pi,0)$.

The most remarkable results here are the continuous versus abrupt
changes in ARPES and transport studies and, respectively, in the Raman spectra across the
critical doping level of $p \approx 0.21$. Apparently, we encounter
a transition from an essentially conventional metallic state to
one of strongly interacting electrons. The origin of the dichotomy
between single- vs two-particle properties has to remain open at the
moment. From the Raman scattering point of view a doping-dependent elastic term can be excluded as an explanation.

%%%%%%%%%%%%%%%%%%%%%%%%%%%%%%%%%%%%%%%%%%%%%%%%%%%
\begin{figure*}
  \centering
  \resizebox{1.\columnwidth}{!}{\includegraphics{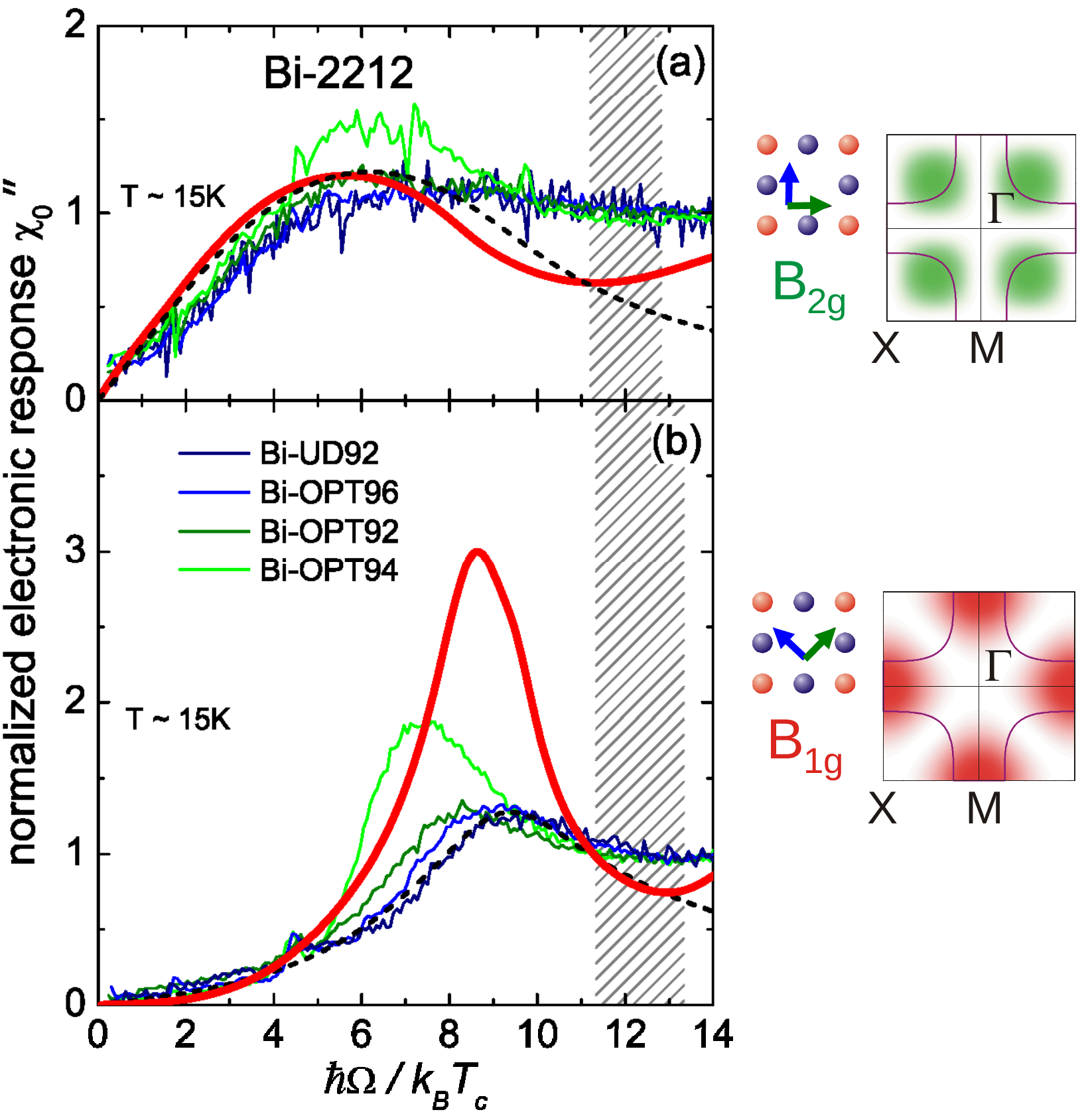}}
  \caption{Electronic Raman response of $\rm Bi_2Sr_2CaCu_2O_{8+\delta}$
  well below $T_c$ close to optimal doping. The
  energy is given in units of $k_BT_c$. The hatched area
  indicates the energy range, where superconducting and normal state
  spectra merge. The dashed and the full smooth (red) line represents the weak and strong coupling predictions, respectively, for the superconducting gap structures.
  }
  \label{fig:sc}
\end{figure*}
%%%%%%%%%%%%%%%%%%%%%%%%%%%%%%%%%%%%%%%%%%%%%%%%%%%%%%%%%%%%%%%%%%%%%%%%%%%%%%%%%%

Below $T_c$ an additional complication arises. While the $B_{2g}$ response [Fig.~\ref{fig:sc}~(a)] is universal, the $B_{1g}$ spectra clearly reflect sample specific behavior and the
superconductivity-induced features exhibit a sta\-tistically
significant sample dependence at a given doping level as shown in Fig.~\ref{fig:sc}~(b). More quantitatively, the $B_{2g}$ response [Fig.~\ref{fig:sc}~(a)] scales with the individual $T_c$, and is  satisfactorily described already by the weak-coupling prediction. There are only small differences in the intensity close to the peak maximum which can be traced back to variations in the impurity concentration \cite{Devereaux:1995}. In contrast, the $B_{1g}$ spectra [Fig.~\ref{fig:sc}~(b)] clearly reflect sample
specific behavior and, beyond that, do not scale with $T_c$ but
rather as $(1-p)$ \cite{Devereaux:2007,Kendziora:1995,Nemetschek:1997,Sugai:2000,Opel:2000,Venturini:2002,Hackl:2005,LeTacon:2006,Munnikes:2009}.
Moreover, on the basis of results obtained with applied pressure
\cite{Goncharov:2003} we conclude that internal strain induced by
quenched disorder \cite{Eisaki:2004} leads to the variation of the
$B_{1g}$ spectra below $T_c$.  This experimental fact escaped attention so far
but may shed light on the origin of the $B_{1g}$ spectra. It is
indeed hard to understand their origin in terms of a pure
pair-breaking effect.

This qualitative reasoning is fully corroborated by the
simulations derived from the ARPES spectra. Here,
Eq.~(\ref{eq:chi}) must be formulated for superconductivity, and the diagonal (normal) and the off-diagonal (anomalous) parts of the Green function \cite{Devereaux:2007} are needed. The self energies have to be derived in the spirit of a
strong coupling approach
\cite{Cuk:2005,Carbotte:1999,Eschrig:2003,Inosov:2007}. The agreement with the $B_{2g}$ spectra is comparable to that of the weak-coupling prediction (Fig.~\ref{fig:sc}~(a)). However, there is no agreement between the simulations and the $B_{1g}$ Raman spectra [Fig.~\ref{fig:sc}~(b)] although sample
Bi-UD92 came from the same source as the one used by Inosov \textit{et al.} \cite{Inosov:2007}
for ARPES. Results for samples from other sources lie
more or less inside the pair-breaking peak derived from ARPES but,
similarly as in the normal state, the overall experimental intensity is
generally too small. The larger intensity may originate in the Fermi liquid like variation of the self energy proposed by Inosov \textit{et al.} \cite{Inosov:2007} to fit the ARPES spectra which leads also to discrepancies in $B_{2g}$ symmetry above the maximum [Fig.~\ref{fig:sc}~(a)]. At lower energies $\Sigma^{\prime\prime}$ appears to vary closer to linearly \cite{Cooper:2009,Devereaux:1995} rather than quadratically.

\section{Conclusions}
\label{sec:4}
The quantitative comparison of  ARPES and Raman spectra in lowest
order approximation
yields quantitative agreement of the normal-state response above $p \approx 0.21$. The different spectral shapes of the $B_{1g}$ and $B_{2g}$ symmetry are obtained with a momentum independent self energy and can be traced back to the band structure alone, specifically to the proximity of the van Hove singularity to the chemical potential. At $p \le 0.21$, discrepancies between the single- and two-particle responses are observed predominantly in $B_{1g}$
symmetry while nodal ARPES and $B_{2g}$ Raman spectra remain consistent by and large. The $B_{1g}$ Raman spectra in both the normal state and the superconductivity-induced features are progressively suppressed in a frequency range of at least 1500\,cm$^{-1}$. As opposed to transport measurements \cite{Nakamae:2003,Cooper:2009}, a momentum dependent constant relaxation term alone which, for instance, could originate from correlated scattering centers \cite{Kampf:1997} is insufficient to explain the differences. We rather conclude, that the origin of the discrepancies is due to dynamic processes. The abrupt onset of  renormalization effects in the
$B_{1g}$ spectra suggests a crossover controlled by a quantum critical point close to the center
of the superconducting dome \cite{Venturini:2002b}. A new type of interaction appears which is not fully described by the single-particle self energy. It is an interesting question as to whether or not the suppression of spectral weight has its origin in the the same interactions as those which favor a new ground state and induce superconductivity.

Similarly as in the normal state, the $B_{2g}$  responses below $T_c$ look rather conventional and scale with the individual transition temperatures in the entire doping range studied  \cite{Munnikes:2009}. The explanation in terms of a weak-coupling approach is limited to $\Omega < 2\Delta_0$. In a strong coupling approximation with the self energy and the band structure derived from the ARPES results an energy range of at least a few $\Delta_0$ can be assessed. This is not the case in $B_{1g}$
symmetry. At optimal doping, $p=0.16$, the pair-breaking
features observed by Raman scattering reflect properties of the
individual samples and are inconsistent with the single-particle
results. This implies that the  $B_{1g}$ spectra do not directly and exclusively reflect the maximal gap $\Delta_0$. Rather, pairing correlations induce features close to $2\Delta_0$ and, additionally, seem to activate another excitation which is particularly clearly seen for sample Bi-OPT94. In general, the $B_{1g}$ response appears to provide crucial yet unexplained information for the understanding of the relevant interactions in the cuprates.

\section*{Acknowledgements}
We are indebted to L. Benfatto, S. Caprara, T.P. Devereaux, C. Di~Castro, M. Grilli, B. Moritz, and A. Zawadowski for valuable discussions. The work has been supported by the DFG via Research Unit FOR~538 (grant-no. Ha2071/3). B.M. and R.H. gratefully acknowledge support by the Bavarian Californian Technology Center (BaCaTeC).

%%%%%%%%%%%%%%%%%%%%%%%%%%%%%%%%%%%%%%%%%%%%%%%%%%%%%%%%%%%%%%%%%%

\end{document}